\documentclass[10pt, conference, letterpaper]{IEEEtran}
\IEEEoverridecommandlockouts
\usepackage{cite}
\usepackage{amsmath,amssymb,amsthm}
\usepackage{bbm}
\usepackage{algpseudocode}
\usepackage{graphicx}
\usepackage{textcomp}
\usepackage{mathtools}
\usepackage{xcolor}
\usepackage{algorithm}
\usepackage{bm}
\usepackage{booktabs,threeparttable}

\algnewcommand\algorithmicswitch{\textbf{switch}}
\algnewcommand\algorithmiccase{\textbf{case}}
\algdef{SE}[SWITCH]{Switch}{EndSwitch}[1]{\algorithmicswitch\ #1\ \algorithmicdo}{\algorithmicend\ \algorithmicswitch}%
\algdef{SE}[CASE]{Case}{EndCase}[1]{\algorithmiccase\ #1}{\algorithmicend\ \algorithmiccase}%
\algtext*{EndSwitch}%
\algtext*{EndCase}%

\def\BibTeX{{\rm B\kern-.05em{\sc i\kern-.025em b}\kern-.08em
    T\kern-.1667em\lower.7ex\hbox{E}\kern-.125emX}}
\usepackage{url}

\usepackage{tikz}
\usetikzlibrary{shapes, arrows, positioning,plotmarks,patterns}

\usepackage{subfig}

\usepackage{pgfplots}
\pgfplotsset{
compat=1.3,
legend style={fill opacity=0.7,  draw opacity=1, text opacity=1, draw=white!15!black, legend cell align=left, align=left},
width=6cm, 
height=6cm,
yminorticks=false,
xminorticks=false,
title style={font=\small},
tick style={color=black},
tick label style={font=\small},
grid style={line width=.1pt, draw=gray!20},
major grid style={line width=.1pt,draw=gray!20},
}
\usepgfplotslibrary{colorbrewer}
\usepgfplotslibrary{groupplots}
\usepgfplotslibrary{fillbetween}
\usetikzlibrary{patterns}
\pgfplotsset{every tick label/.append style={font=\footnotesize}}

\usepackage{glossaries}
\newacronym{zw}{ZW}{Zero-Wait}
\newacronym{cr}{CR}{Collision Resolution}
\newacronym{ce}{CE}{Collision Exit}
\newacronym{bt}{BT}{Belief Threshold}
\newacronym{aoii}{AoII}{Age of Incorrect Information}
\newacronym{paoii}{PAoII}{Peak AoII}
\newacronym{aoi}{AoI}{Age of Information}
\newacronym{cdf}{CDF}{Cumulative Distribution Function}
\newacronym{pmf}{PMF}{Probability Mass Function}
\newacronym{delta}{DELTA}{Dynamic Epistemic Logic for Tracking Anomalies}
\newacronym{del}{DEL}{Dynamic Epistemic Logic}
\newacronym{tdd}{TDD}{Time Division Duplex}
\newacronym{sack}{SACK}{Selective Acknowledgment}
\newacronym{ack}{ACK}{acknowledgment}

\newacronym{rr}{RR}{Round-Robin}
\newacronym{maf}{MAF}{Maximum Age First}
\newacronym{lzw}{LZW}{Local Zero-Wait}
\newacronym{gzw}{GZW}{Global Zero-Wait}
\newacronym{fs}{FS}{Frame Slotted}

\DeclareMathOperator*{\Prob}{Pr}    
\newcommand{\Prb}[1]{\Prob\left[ #1 \right]} 
\newcommand{\mc}[1]{\mathcal{#1}}   
\newcommand{\mb}[1]{\mathbf{#1}}    
\DeclareMathOperator{\Bin}{Bin}    

\algtext*{EndWhile}
\algtext*{EndIf}
\algtext*{EndFor}

\def \sfwidth{0.32\linewidth}
\def \sfheight {0.21\linewidth}
\def \boxside {0.12\linewidth}

\definecolor{color0}{HTML}{00429D}
\definecolor{color1}{HTML}{844D99}
\definecolor{color2}{HTML}{C3608E}
\definecolor{color3}{HTML}{EF8078}
\definecolor{color4}{HTML}{FFB047}
\definecolor{darkslategray38}{RGB}{38,38,38}

\makeatletter

\title{Peak Age of Incorrect Information of Reactive ALOHA Reporting Under Imperfect Feedback}

\author{\IEEEauthorblockN{Federico Chiariotti\IEEEauthorrefmark{1}, Andrea Munari\IEEEauthorrefmark{2}, Leonardo Badia\IEEEauthorrefmark{1}, and Petar Popovski\IEEEauthorrefmark{3}}
\IEEEauthorblockA{\IEEEauthorrefmark{1}Department of Information Engineering, University of Padova, 35131 Padua, Italy\\
\IEEEauthorrefmark{2}Institute of Communications and Navigation, German Aerospace Center (DLR), We{\ss}ling, Germany\\
\IEEEauthorrefmark{3}Department of Electronic Systems, Aalborg University, 9220 Aalborg, Denmark\\
Emails: \small{\texttt{chiariot@dei.unipd.it}, \texttt{andrea.munari@dlr.de}, \texttt{leonardo.badia@unipd.it}, \texttt{petarp@es.aau.dk}}}
}

\begin{document}

\maketitle

\begin{abstract}
\gls{aoii} is particularly relevant in systems where real time responses to anomalies are required, such as natural disaster alerts, cybersecurity warnings, or medical emergency notifications. Keeping system control with wrong information for too long can lead to inappropriate responses.
In this paper, we study the \gls{paoii} for multi-source status reporting by independent devices over a collision channel, following a zero-threshold ALOHA access where nodes observing an anomaly immediately start transmitting about it. If a collision occurs, nodes reduce the transmission probability to allow for a resolution. Finally, wrong or lost feedback messages may lead a node that successfully updated the destination to believe a collision happened. The \gls{paoii} for this scenario is computed in closed-form. We are eventually able to derive interesting results concerning the minimization of \gls{paoii}, which can be traded against the overall goodput and energy efficiency, but may push the system to the edge of congestion collapse.
\end{abstract}


\glsresetall

\section{Introduction}\label{sec:intro}

Modern communication systems provide a plethora of real-time services in which timely and accurate information is critical~\cite{bedewy2021optimal}. This is especially true for alert systems designed to notify users of events or conditions requiring immediate action~\cite{abreu2020scheduling}. Over the last decade, \gls{aoi} has emerged as a practical metric to quantify timeliness of information~\cite{yates2021age,kosta2017age}. However, alongside \gls{aoi}, the accuracy of the notifications and a timely tracking of critical status changes in the system play an equally pivotal role, leading to the popularity of the more recently proposed \gls{aoii}~\cite{maatouk2020age}.

In scenarios involving safety, health, or tactical monitoring, obsolete information can still be useful as long as it reflects the system condition, whereas failure to timely report that the last status report has become incorrect can have catastrophic consequences. We argue that in these systems a suitable goal metric to minimize would be, rather than \gls{aoi} alone, the value of \gls{paoii}, i.e., the maximum duration for which system control is exposed to incorrect or misleading data before a correction is received. This metric is valuable for emergency notifications and whenever keeping erroneous data for too long can lead to inappropriate responses.

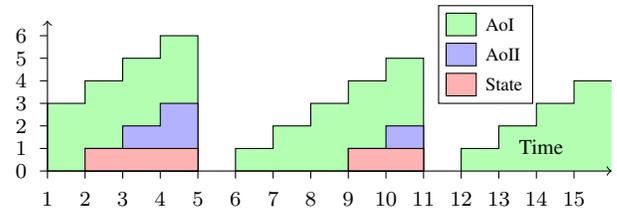
\begin{figure}[t!]
    \centering
    \begin{tikzpicture}

\path [fill=green!30!white,draw] (0,1.9)--(0.5,1.9)--(0.5,2.2)--(1,2.2)--(1,2.5)--(1.5,2.5)--(1.5,2.8)--(2,2.8)--(2,1)--(0,1)--cycle;

\path [fill=green!30!white,draw] (2.5,1.3)--(3,1.3)--(3,1.6)--(3.5,1.6)--(3.5,1.9)--(4,1.9)--(4,2.2)--(4.5,2.2)--(4.5,2.5)--(5,2.5)--(5,1)--(2.5,1)--cycle;

\path [fill=green!30!white] (5.5,1.3)--(6,1.3)--(6,1.6)--(6.5,1.6)--(6.5,1.9)--(7,1.9)--(7,2.2)--(7.5,2.2)--(7.5,1)--(5.5,1)--cycle;

\path [fill=blue!30!white,draw] (0.5,1)--(0.5,1.3)--(1,1.3)--(1,1.6)--(1.5,1.6)--(1.5,1.9)--(2,1.9)--(2,1)--cycle;

\path [fill=blue!30!white,draw] (4,1)--(4,1.3)--(4.5,1.3)--(4.5,1.6)--(5,1.6)--(5,1)--cycle;

\path [fill=red!30!white,draw] (0.5,1)--(0.5,1.3)--(2,1.3)--(2,1)--cycle;

\path [fill=red!30!white,draw] (4,1)--(4,1.3)--(5,1.3)--(5,1)--cycle;

\path [draw] (5.5,1)--(5.5,1.3)--(6,1.3)--(6,1.6)--(6.5,1.6)--(6.5,1.9)--(7,1.9)--(7,2.2)--(7.5,2.2);

\draw[->] (0,1) --  node [above=0.1cm,very near end]{\footnotesize Time} (7.5,1);
\draw[->] (0,1) -- (0,3);
\foreach \i in {1,...,15}
    \draw[-] (0.5*\i-0.5,0.9)  --node[below=0.1cm]{\footnotesize$\i$} (0.5*\i-0.5,1);
\foreach \i in {0,...,6}
    \draw[-] (-0.1,0.3*\i+1) --node[left=0.1cm]{\footnotesize$\i$} (0,0.3*\i+1);

\path [fill=green!30!white,draw] (5.3,3.1)--(5.3,2.8)--(5.7,2.8)--(5.7,3.1)--cycle;

\path [fill=blue!30!white,draw] (5.3,2.4)--(5.3,2.7)--(5.7,2.7)--(5.7,2.4)--cycle;

\path [fill=red!30!white,draw] (5.3,2)--(5.3,2.3)--(5.7,2.3)--(5.7,2)--cycle;

\path [draw] (5.2,1.9)--(5.2,3.2)--(6.45,3.2)--(6.45,1.9)--cycle;

\node[anchor=west] at (5.7,2.15){\scriptsize State};
\node[anchor=west] at (5.7,2.55){\scriptsize AoII};
\node[anchor=west] at (5.7,2.95){\scriptsize AoI};

\end{tikzpicture}
    \caption{Example of the AoI and AoII evolution for a node.}\vspace{-0.3cm}
    \label{fig:aoii_diagram}
\end{figure}

In this paper, we analyze \gls{paoii} for a network of users, each reporting over different sensed quantities to a common sink. The information content of each sensor is not correlated with the others and the access procedures are also distributed, following an ALOHA model over a collision channel~\cite{ma2008analysis}. A seminal analysis of \gls{aoi} for ALOHA systems is already presented in~\cite{kaul2011minimizing}, and has seen many follow-up extensions involving carrier-sense~\cite{maatouk2020ageCSMA}, capture effect~\cite{badia2024game}, or finite horizon scheduling~\cite{hegde2024age}. \gls{fs} ALOHA~\cite{okada1977analysis} has also been shown to be particularly convenient for \gls{aoi} minimization over multi-access channels due to its framed structure that decreases the variance of the individual success probability. In~\cite{yue2023age}, an analysis of \gls{fs} ALOHA was performed, focusing only on \gls{aoi}. Moreover,~\cite{yavascan2021analysis} analyzed \gls{aoi} in slotted ALOHA by considering an age threshold, which is shown to benefit information freshness by decreasing collisions and allowing nodes with staler information to gain priority. However, this approach is conceived without considering \gls{aoii} and status inaccuracy. A threshold-based approach and a framed structure are combined in~\cite{wang2025age}, which applies a threshold mechanism to \gls{fs} ALOHA, with the same goal of improving \gls{aoi}.
Protocol enhancements leveraging decentralized scheduling to improve \gls{aoi} have been also presented in~\cite{chen2022age}, essentially based on the degree of innovation that packets bring, and in~\cite{jiang2018decentralized}, which establishes a round robin schedule among competing users.

Fig.~\ref{fig:aoii_diagram} highlights the significant difference between traditional \gls{aoi} and \gls{aoii}. The latter is suited to anomaly reporting applications, where the interest in the freshness of the state depends on the state, resulting in protocols that operate very differently as compared to the ones optimized for \gls{aoi}. The studies that optimize protocols for \gls{aoii} are rather sporadic. \gls{aoii} is evaluated with a Markov model in~\cite{shao2023average}, for both slotted and \gls{fs} ALOHA versions. In~\cite{nayak2023decentralized}, the goal of \gls{aoii} minimization is tackled with a distributed approach, and finally~\cite{munari2024monitoring} compares different strategies for update reporting over a multi-access collision channel. In this work, we combine the novelty of \gls{aoii} analysis with an often-neglected aspect, i.e., the reliability of the feedback channel.

Our study can be seen as considering the \emph{reactive} strategy, where transmissions are made by sensing nodes only when a change in their observed metric occurs, but without waiting for any threshold, until they eventually reach acknowledgment from the sink. The effect of lost feedback messages~\cite{badia2009effect} causes additional unnecessary transmissions from nodes mistakenly believing that an update was lost. In other words, each node is in an idle state until it observes an anomaly, in which case it starts reporting about it until the message is eventually received at the common sink. If collisions occur, they cause the nodes to enter a resolution phase with lower access probability. However, failure to receive an \gls{ack} may erroneously lead a node to believe that a collision occurred. We analyze the system and optimize the protocol parameters, with the interesting conclusion that a value precisely minimizing the \gls{paoii} can be found on the brink of system instability, but preceded by a general plateau for most other key performance indicators such as goodput and power consumption. Therefore, the worst-case \gls{paoii} can be optimized, possibly leaving a proper margin to avoid instability.


\section{System Model}\label{sec:system}

We consider a scenario in which a set $\mc{N}$ of $N=|\mc{N}|$ sensors, need to report anomalies to a single gateway over a shared 
channel. We consider a slotted time system; at each time step, an anomaly may be independently detected by any sensor with probability $\lambda$. Anomalies persist until they are successfully reported to the gateway. The state of the measured process at time $t$ is $\mb{x}(t)\in\{0,1\}^N$: if $x_n(t)=1$, there is an unreported anomaly for sensor $n$. The Markovian transition matrix $\mb{H}_n$ for the state $x_n(t)$ of sensor $n$ is then
\begin{equation}
\mb{H}_n=\begin{pmatrix}
    1-\lambda & \lambda\\
    \xi_{n}(t) & 1-\xi_{n}(t)
\end{pmatrix},
\end{equation}
where $\xi_{n}(t)$ is an indicator variable, equal to $1$ if $n$ successfully transmits a packet in slot $t$ and $0$ otherwise. Following the common definition \cite{maatouk2020age}, the \gls{aoii} $\Delta_n(t)$ is
\begin{equation}
  \Delta_n(t)=\mathbb{I}\left(x_n(t)=1\right)\Delta_n(t-1)+1,
\end{equation}
where $\mathbb{I}(\cdot)$ is the indicator function, equal to $1$ if the argument is true and $0$ otherwise.
The \gls{aoii} grows only when the sensor is in state $1$, i.e., there is an unreported anomaly, and resets to $0$ when the node successfully transmits an update. We also define the \gls{paoii} $\theta_n$, i.e., the maximum \gls{aoii} reached before the anomaly is successfully reported. This is equivalent to sampling the \gls{aoii} when $x_n(t)\xi_n(t)=1$.

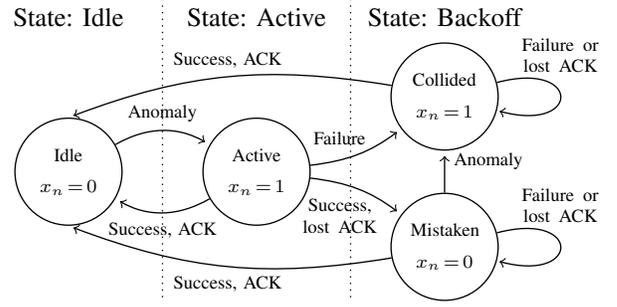
\begin{figure}[t!]
    \centering
    \begin{tikzpicture}[->, shorten >=2pt, line width=0.5 pt, node distance =1 cm]
                        ]

\node (idle) [circle,draw,minimum height=1cm, text width=0.9cm,align=center] at (0,0) {\scriptsize Idle $x_n\!=\!0$};
\node (active) [circle,draw,minimum height=1cm, text width=0.9cm,align=center] at (2.5,0) {\scriptsize Active $x_n\!=\!1$};
\node (backoff) [circle,draw,minimum height=1cm, text width=0.9cm,align=center] at (5,1) {\scriptsize Collided $x_n\!=\!1$};
\node (mistaken) [circle,draw,minimum height=1cm, text width=0.9cm,align=center] at (5,-1) {\scriptsize Mistaken $x_n\!=\!0$};

\path (idle) edge  [bend left] node [above]{\scriptsize Anomaly} (active);
\path (active) edge  [bend left] node [below]{\scriptsize Success, ACK} (idle);

\path (active) edge [bend left=15] node [below,xshift=-0.25cm,text width=1.1cm, align=center,execute at begin node=\setlength{\baselineskip}{8pt}]{\scriptsize Success, lost ACK} (mistaken);

\path (active) edge [bend right=15] node [above,xshift=-0.25cm,text width=1.1cm, align=center]{\scriptsize Failure} (backoff);

\path (mistaken) edge node [right,near end]{\scriptsize Anomaly} (backoff);

\path (mistaken) edge [bend left=15] node [below]{\scriptsize Success, ACK} (idle.south);

\path (backoff) edge [bend right=15] node [above]{\scriptsize Success, ACK} (idle.north);

\path (mistaken) edge [loop right] node [above,text width=1.2cm,yshift=0.2cm, align=center,execute at begin node=\setlength{\baselineskip}{8pt}] {\scriptsize Failure or lost ACK} (mistaken);

\path (backoff) edge [loop right] node [above,text width=1.2cm,yshift=0.2cm, align=center,execute at begin node=\setlength{\baselineskip}{8pt}] {\scriptsize Failure or lost ACK} (backoff);

\node at (0,2.05){State: Idle};
\node at (2.5,2.05){State: Active};
\node at (5,2.05){State: Backoff};

\draw[-,dotted] (1.25,2.3) -- (1.25,-1.8);
\draw[-,dotted] (3.75,2.3) -- (3.75,-1.8);

\end{tikzpicture}
    \caption{Protocol state diagram.}\vspace{-0.3cm}
    \label{fig:protocol_state}
\end{figure}

Our communication model is a 
collision channel: if more than one sensor attempts a transmission in the same slot, no packet is received. On the other hand, there is an error probability $\varepsilon$ even if a single sensor transmits, representing the channel uncertainty. Given vector $\bm{\gamma}(t)$, with $\gamma_n(t)=1$ if sensor $n$ attempts a transmission in the slot and $0$ otherwise, the probability of success is then
\begin{equation}
  \Prb{\textstyle\sum_{n\in\mc{N}}\xi_n(t)=1|\bm{\gamma}(t)}=\mathbb{I}\left(||\bm{\gamma}(t)||=1\right)(1-\varepsilon),
\end{equation}
where $||\cdot||$ is the norm of a vector. If a sensor is successful, we consider a latency of $1$ slot, i.e., the \gls{paoii} if the sensor transmits immediately is equal to $1$. Furthermore, we consider an imperfect feedback channel operating over \gls{tdd}: at the end of each slot, the gateway transmits a short \gls{ack} packet. Transmitting nodes then switch to receiving mode, but the \gls{ack} may be lost with probability $\psi$.

We then consider a slotted ALOHA protocol with backoff, in which nodes may be in $3$ different states:
  (i)
  \emph{Idle} nodes have nothing to transmit;
  (ii)
  \emph{Active} nodes need to transmit an anomaly, and attempt a transmission in each slot with probability $\alpha$;
  (iii)
  \emph{Backoff} nodes did not receive an \gls{ack}, and thus switch to the backoff transmission probability $\beta$.

The protocol state diagram is represented in Fig.~\ref{fig:protocol_state}, in which we distinguish between \emph{collided} nodes, i.e., sensors that enter backoff mode after a packet is lost in the uplink phase, and \emph{mistaken} nodes, whose packet was received correctly but which missed the relative \gls{ack}. From the perspective of the sensor, these events are completely indistinguishable, but the distinction is useful in the analysis. Mistaken nodes may become collided if a new anomaly appears.

\section{Protocol Analysis}\label{sec:analysis}

We can now analyze the system as a Markov chain, considering a compact state $\mb{s}=\langle a,c,m\rangle$, which represent the number of active, collided, and mistaken sensors, respectively.
Naturally, $a+c+m\leq N$, and we define the auxiliary variable $I=N-a-c-m$, i.e., the number of idle sensors. We denote the set of possible states as $\mc{S}$ and the change in $a$ when transiting from state $\mb{s}$ to state $\mb{s}'$ as $d_a=a'-a$, and define $d_c$ and $d_m$ analogously.

\subsection{Steady-State Analysis}

First, we can easily compute the probability that $i$ idle sensors and $m$ mistaken ones will detect a new anomaly as
\begin{equation}
  \zeta_{\mb{s}}^{(i,j)}=\Bin^{I}_{\lambda}(i)\Bin^{m}_{\lambda}(j),
\end{equation}
where $\Bin^N_p(m)$ is the binomial \gls{pmf}. On the other hand, the probabilities of success for a specific node in active and backoff mode when there are $a$ nodes in active mode and $b$ nodes in backoff mode are
\begin{align}
  \sigma_a(a,b)=&(1-\varepsilon)\alpha(1-\alpha)^{a-1}(1-\beta)^{b};\\
  \sigma_b(a,b)=&(1-\varepsilon)(1-\alpha)^a\beta(1-\beta)^{b-1}.
\end{align}

We can then distinguish a number of cases, listed in Table~\ref{tab:cases}.\footnote{Noise-induced errors on both channels, as well as collisions, are directly captured by the model. However, the current version does not consider the possibility of packet capture: this could be easily introduced as a further division of case $\mc{H}$, and does not significantly affect the results.} In particular, if the system is in state $\mb{s}=\langle a,c,m\rangle$ and $i\in\{0,\ldots,I\}$ idle nodes and $j\in\{0,\ldots,m\}$ mistaken nodes detect a new anomaly, we have these transitions from $\mb{s}$ to $\mb{s}'$:
\begin{itemize}
    \item \emph{Case $\mc{A}$}: An active node successfully transmits and gets the \gls{ack}, $\mb{s}'=\langle a+i-1,c+j,m-j\rangle$.
    \item \emph{Case $\mc{B}$}: An active node successfully transmits, but the \gls{ack} is lost, $\mb{s}'=\langle a+i-1,c+j,m-j+1\rangle$.
    \item \emph{Case $\mc{C}$}: A collided node successfully transmits and gets the \gls{ack}, $\mb{s}'=\langle a+i,c+j-1,m-j\rangle$.
    \item \emph{Case $\mc{D}$}: A collided node successfully transmits, but the \gls{ack} is lost, $\mb{s}'=\langle a+i,c+j-1,m-j+1\rangle$.
    \item \emph{Case $\mc{E}$}: A mistaken node successfully transmits and gets the \gls{ack}, $\mb{s}'=\langle a+i,c+j,m-j-1\rangle$.
    \item \emph{Case $\mc{F}$}: A mistaken node successfully transmits, but the \gls{ack} is lost, $\mb{s}'=\langle a+i,c+j,m-j\rangle$.
    \item \emph{Case $\mc{G}$}: No transmission, $\mb{s}'=\langle a+i,c+j,m-j\rangle$.
    \item \emph{Case $\mc{H}$}: Collision involving $k\in\{0,\ldots,a+i\}$ active nodes, $\mb{s}'=\langle a+i-k,c+j+k,m-j\rangle$.
    \item \emph{Case $\mc{I}$}: An active node is the only one to transmit, but fails, $\mb{s'}=\langle a+i,c+j,m-j\rangle$.
    \item \emph{Case $\mc{J}$}: A node in backoff mode (either mistakenly or correctly) transmits and fails, $\mb{s}'=\langle a+i,c+j,m-j\rangle$.
\end{itemize}
We can then compute the probability of each case:
\begin{equation}
    p_{i,j}^{\kappa}=\begin{cases}
        (1-\psi)(a+i)\sigma_a(a\!+\!i,c\!+\!m)\zeta_{\mb{s}}^{(i,j)}, &\kappa=\mc{A};\\
        \psi(a+i)\sigma_a(a\!+\!i,c\!+\!m)\zeta_{\mb{s}}^{(i,j)}, &\kappa=\mc{B};\\
        (1-\psi)(c+j)\sigma_b(a\!+\!i,c\!+\!m)\zeta_{\mb{s}}^{(i,j)}, &\kappa=\mc{C};\\
        \psi(c+j)\sigma_b(a\!+\!i,c\!+\!m)\zeta_{\mb{s}}^{(i,j)}, &\kappa=\mc{D};\\
        (1-\psi)(m-j)\sigma_b(a\!+\!i,c\!+\!m)\zeta_{\mb{s}}^{(i,j)}, &\kappa=\mc{E};\\
        \psi(m-j)\sigma_b(a\!+\!i,c\!+\!m)\zeta_{\mb{s}}^{(i,j)}, &\kappa=\mc{F};\\
        (1-\alpha)^{a+i}(1-\beta)^{c+m}\zeta_{\mb{s}}^{(i,j)}, &\kappa={G};\\
        \varepsilon\Bin_{\alpha}^{a+i}(1)\Bin_{\beta}^{c+m}(0)\zeta_{\mb{s}}^{(i,j)}, &\kappa=\mc{I};\\
        \varepsilon\Bin_{\alpha}^{a+i}(0)\Bin_{\beta}^{c+m}(1)\zeta_{\mb{s}}^{(i,j)}, &\kappa=\mc{J}.
    \end{cases}
\end{equation}
We consider case $\mc{H}$ separately, as we need to keep track of the number $k$ of active nodes involved in the collision:
\begin{equation}
  p_{i,j,k}^{\mc{H}}(\mb{s})=\Bin_{\alpha}^{a+i}(k)\sum_{\mathclap{\ell=\max(0,2-k)}}^{c+m}\Bin_{\beta}^{c+m}(\ell)\zeta_{\mb{s}}^{(i,j)}.
\end{equation}

Finally, we combine these events to obtain the non-zero state transition probabilities. Each element of the transition matrix $\mb{M}$ is then defined as
\begin{equation}
M_{\mb{s},\mb{s}'}=
\begin{cases}
p_{d_{a},d_c\!+\!1}^{\mc{C}}(\mb{s})\!+\!p_{d_{a},d_{c}}^{\mc{E}}(\mb{s}),&d_{c}\!=\!-\!1\!-\!d_m;\\
\begin{aligned}
    &p_{d_a\!+\!1,d_{c}}^{\mc{A}}\!(\mb{s})\!+\!\sum\limits_{\mathclap{\kappa\in\{\mc{F},\mc{G},\mc{J}\}}}p_{d_{a},d_{c}}^{\kappa}(\mb{s})\\
    &+p_{d_{a},d_{c},0}^{\mc{H}}(\mb{s})\!+\!p_{d_{a},d_c\!+\!1}^{\mc{D}}(\mb{s}),
\end{aligned}
&d_c\!=\!-\!d_m;\\
\begin{aligned}
    &p_{d_a+1,d_{c}}^{\mc{B}}(\mb{s})+p_{d_a+1,d_c-1}^{\mc{I}}(\mb{s})\\&+p_{d_a+1,d_c-1,1}^{\mc{H}}(\mb{s}),\end{aligned}
&d_{c}\!=\!1\!-\!d_{m};\\
    p_{d_{a}+d_{c}+d_{m},-d_{m},d_{c}+d_{m}}^{\mc{H}}(\mb{s}), &d_{c}\!>\!1\!-\!d_{m}.
\end{cases}
\end{equation}
Naturally, the events only have a non-zero probability if their parameters are within the feasible bounds. We then use the eigenvalue method to obtain the steady-state distribution $\bm{\pi}$.

\begin{figure}[t!]\vspace{-0.3cm}
\begin{table}[H]
\centering
	\caption{Events and state changes.}
	\label{tab:cases}
    \scriptsize
	\begin{tabular}[c]{c|c|ccc}
		\toprule
		Event & Meaning & $d_{a}$& $d_{c}$ &$d_{m}$\\
		\midrule
        $\mc{A}(i,j)$ & Success from active node &$i-1$ & $j$ & $-j$\\
        $\mc{B}(i,j)$ & Lost \gls{ack} from active node & $i-1$ & $j$ & $-j+1$\\
        $\mc{C}(i,j)$ & Success from collided node & $i$ & $j-1$ & $-j$\\
        $\mc{D}(i,j)$ & Lost \gls{ack} from collided node & $i$ & $j-1$ & $-j-1$\\
        $\mc{E}(i,j)$ & Success from mistaken node & $i$ & $j$ & $-j+1$\\
        $\mc{F}(i,j)$ & Lost \gls{ack} from mistaken node & $i$ & $j$ & $-j$\\
        $\mc{G}(i,j)$ & Silence & $i$ & $j$ & $-j$\\
        $\mc{H}(i,j,k)$ & Collision & $i-k$ & $j+k$ & $-j$\\
        $\mc{I}(i,j)$ & Loss from active node & $i-1$ & $j+1$ & $-j$\\
        $\mc{J}(i,j)$ & Loss from backoff node & $i$ & $j$ & $-j$\\
		\bottomrule
	\end{tabular}
\end{table}\vspace{-0.6cm}
\end{figure}

\subsection{Goodput, Energy, and Peak AoII Analysis}

The goodput $G$ of the system, defined as the average number of novel packets that are transmitted per slot, can be computed as the steady-state probability of events $\mc{A}$, $\mc{B}$, $\mc{C}$, and $\mc{D}$:
\begin{equation}
  G=\sum_{\mathclap{\mb{s}\in\mc{S}}}\pi(\mb{s})\sum_{i=0}^{I}\sum_{j=0}^m \sum_{\xi\in\{\mc{A},\mc{B},\mc{C},\mc{D}\}} p_{i,j}^{\xi}(\mb{s}).
\end{equation}
In cases $\mc{E}$ and $\mc{F}$, the successfully transmitted information is stale, as it comes from mistaken nodes. In the same way, we can compute the expected power consumption of each node, assuming each slot in which a sensor is active requires energy $E$, including both the packet transmission and \gls{ack} reception, and the slot duration is $\tau$:
\begin{equation}
P=\frac{E}{N\tau}\sum_{s\in\mc{S}}\pi(\mb{s})\left[(c+m)\beta+\sum_{i=0}^I(a+i)\alpha\right].
\end{equation}

We can now compute the \gls{pmf} of the \gls{paoii}. New anomalies may be detected by idle and mistaken nodes, so the probability of a new anomaly being detected in state $\mb{s}$ can be computed by applying Bayes' theorem:
\begin{equation}
  p_{\text{pkt}}(\mb{s})=\frac{\pi(\mb{s})(m+I)\lambda}{\sum_{\mb{s}'\in\mc{S}}\pi(\mb{s}')(I'+m')\lambda}.
\end{equation}
The probabilities are stored as row vector $\mb{p}_{\text{pkt}}$, while column vector $\mb{p}_{\text{id}}$, with $p_{\text{id}}(\mb{s})=\frac{I}{m+I}$, contains the probability that the new anomaly came from an idle node in each state. Finally, we include the success probability for active and backoff nodes in the column vectors $\bm{\eta}^{(t)}$ and $\bm{\nu}^{(t)}$, respectively:
\begin{align}
    \eta^{(t)}(\mb{s})=&\sum_{i=0}^{\mathclap{I-\mathbb{I}(t=1)}}\Bin_{\lambda}^{I-\mathbb{I}(t=1)}(i)\sigma_a(a+i,c+m);\\
    \nu^{(t)}(\mb{s})=&\sum_{i=0}^{\mathclap{I}}\Bin_{\lambda}^{I}(i)\sum_{j=0}^{\mathclap{m-\mathbb{I}(t=1)}}\Bin_{\lambda}^{m-\mathbb{I}(t=1)}(j)\sigma_b(a+i,c+m).
\end{align}
The overall probability of the \gls{paoii} $\theta$ being equal to $1$, i.e., of the new anomaly being immediately reported, can be computed via Bayes' theorem:
\begin{equation}
  p_{\theta}(1)=\mb{p}_{\text{pkt}}\left(\mb{p}_{\text{id}}\odot\bm{\eta}^{(1)}+\left(1-\mb{p}_{\text{id}}\right)\odot\bm{\nu}^{(1)}\right),
\end{equation}
where $\odot$ represents the Hadamard element-wise product.

The following steps are more complex. Nodes that went from idle to active may switch to backoff mode if events $\mc{H}$ or $\mc{I}$ happen. We can then compute the transition matrix in case of an unsuccessful slot for a node in active mode. In the first step, the node goes from idle to active, while in the following steps, it is already active, so there is one less condition on the matrix. If an event results in an active node successfully transmitting or switching to backoff mode, we remove it from the computation:
\begin{IEEEeqnarray}{l}
U^{(t)}_{\mb{s},\mb{s}'}=\\
\begin{cases}
\frac{a'p_{d_a+1,d_{c}}^{\mc{A}}(\mb{s})}{a'+1},
&\begin{aligned}&d_{c}=-\!d_{m}\ \wedge\\
&(d_a\!=\!0\wedge t\!=\!1);\end{aligned}\\
\begin{aligned}
    &\textstyle\frac{a'p_{d_a+1,d_{c}}^{\mc{A}}(\mb{s})}{a'+1}+p_{d_{a},d_c+1}^{\mc{D}}(\mb{s})\\
    &+p_{d_{a},d_{c},0}^{\mc{H}}(\mb{s})+\sum\limits_{\mathclap{\kappa\in\{\mc{F},\mc{G},\mc{J}\}}}p_{d_{a},d_{c}}^{\kappa}(\mb{s}),
    \end{aligned}
    &\begin{aligned}&d_{c}=-d_{m}\ \wedge\\
&(d_a\!>\!0\vee t\!>\!1);&\end{aligned}\\
    p_{d_{a},d_c+1}^{\mc{C}}(\mb{s})+p_{d_{a},d_{c}}^{\mc{E}}(\mb{s}),&\begin{aligned}&d_{c}=-1-d_{m}\ \wedge\\
&(d_a\!>\!0\vee t\!>\!1);\end{aligned}\\
\begin{aligned}&\textstyle\frac{a'}{a'+1}\big[p_{d_a\!+\!1,d_{c}}^{\mc{B}}\!(\mb{s})\!+\!p_{d_a\!+\!1,\!-\!d_m}^{\mc{I}}\!(\mb{s})\\&+\!p_{d_a\!+\!1,\!-\!d_m,1}^{\mc{H}}\!(\mb{s})\big]\end{aligned},&\begin{aligned}&d_{c}=1-d_{m}\ \wedge\\
&(d_a\!>\!0\vee t\!>\!1);\end{aligned}\\
    \frac{a'p_{d_{a}+d_{c}+d_{m},-d_{m},d_{c}+d_{m}}^{\mc{H}}(\mb{s})}{a'+d_{c}+d_{m}}, &\begin{aligned}&d_{c}>1-d_m\ \wedge\\
&(d_a\!+\!d_c\!>\!-d_m\vee t\!>\!1).\end{aligned}
\end{cases}\nonumber
\end{IEEEeqnarray}
In cases in which the node switches to backoff mode, we have
\begin{IEEEeqnarray}{l}
V^{(t)}_{\mb{s},\mb{s}'}=\\
\begin{cases}
    \frac{p_{d_a+1,d_c-1}^{\mc{I}}(\mb{s})+p_{d_a+1,d_c-1,1}^{\mc{H}}(\mb{s})}{a'+1},
&\begin{aligned}&d_{c}=1-d_{m}\ \wedge\\
&(d_a\!\geq\! 0\vee t>1);\end{aligned}\\
    \frac{(d_c+d_m)p_{d_{a}+d_{c}+d_{m},-d_{m},d_{c}+d_{m}}^{\mc{H}}\!(\mb{s})}{a'+d_{c}+d_{m}}, &\begin{aligned}&d_{c}\!+\!d_{m}>1\ \wedge\\
&(d_a\!+\!d_c\!>\!-\!d_m\vee t>1).\end{aligned}
\end{cases} \nonumber
\end{IEEEeqnarray}
Finally, we consider the transition matrix if the considered node is already in backoff mode:
\begin{equation}
\begin{aligned}
&W^{(t)}_{\mb{s},\mb{s}'}=\\
&\begin{cases}
\frac{c'p_{d_{a},d_c+1}^{\mc{D}}(\mb{s})}{c'+1},
&\begin{aligned}&d_{c}=-d_{m}\ \wedge\\&
(d_m=0\wedge t=1);\end{aligned}\\
\begin{aligned}
    &p_{d_a+1,d_{c}}^{\mc{A}}(\mb{s})+\!\sum_{\mathclap{\xi\in\{\mc{F},\mc{G},\mc{J}\}}}p_{d_{a},d_{c}}^{\xi}(\mb{s})\\
    &\!+p_{d_{a},d_{c},0}^{\mc{H}}(\mb{s})+\textstyle\frac{c'p_{d_{a},d_c+1}^{\mc{D}}(\mb{s})}{c'+1},
\end{aligned}&\begin{aligned}&d_{c}=-d_m\ \wedge\\
&(d_m>0\vee t>1);\end{aligned}\\
    \frac{c'p_{d_{a},d_c+1}^{\mc{C}}(\mb{s})}{c'+1},&\begin{aligned}&d_{c}=-1-d_{m}\ \wedge\\
&(d_m=\wedge t=1);\end{aligned}\\[6pt]
    \frac{c'p_{d_{a},d_c+1}^{\mc{C}}(\mb{s})}{c'+1}+p_{d_{a},d_{c}}^{\mc{E}}(\mb{s}),&\begin{aligned}&d_{c}=-1-d_{m}\ \wedge\\
&(d_m>1\vee t>1);\end{aligned}\\
\begin{aligned}
    &p_{d_a+1,d_{c}}^{\mc{B}}(\mb{s})+p_{d_a+1,d_c-1}^{\mc{I}}(\mb{s})\\&+p_{d_a+1,d_c-1,1}^{\mc{H}}(\mb{s}),
\end{aligned}&\begin{aligned}&d_{c}=1-d_{m}\ \wedge\\
&(d_c>0 \vee t>1);\end{aligned}\\
    p_{d_{a}+d_{c}+d_{m},-d_{m},d_{c}+d_{m}}^{\mc{H}}(\mb{s}), &\begin{aligned}&d_{c}>1-d_{m}\ \wedge\\
&(d_m>0\vee t>1).\end{aligned}
\end{cases}
\end{aligned}
\end{equation}

We then define the state posterior probabilities considering active and backoff nodes:
\begin{align}
\mb{p}_{\text{act}}^{(1)}=\mb{p}_{\text{pkt}}\odot\mb{p}_{\text{id}}^T;\quad
\mb{p}_{\text{bk}}^{(1)}=\mb{p}_{\text{pkt}}\odot\left(1-\mb{p}_{\text{id}}\right)^T.
\end{align}
We can compute the state distribution for the following steps if the considered node is in active and backoff mode recursively:
\begin{align}
  \mb{p}_{\text{act}}^{(t+1)}&\!=\!\left(\mb{p}_{\text{act}}^{(t)}\odot\!\left[1\!-\!\bm{\eta}^{(t)}\right]^T\right)\!\mb{U}^{(t)};\\
  \mb{p}_{\text{bk}}^{(t+1)}&\!=\!\left[\mb{p}_{\text{act}}^{(t)}\odot\!\left[1\!-\!\bm{\eta}^{(t)}\right]^T\right]\!\mb{V}^{(t)}\!+\!\left[\!\mb{p}_{\text{bk}}^{(t)}\!\odot\!\left[1\!-\!\bm{\nu}^{(t)}\right]^T\right]\!\mb{W}^{(t)}.
\end{align}
We can then compute the \gls{pmf} of the \gls{paoii} as
\begin{equation}
    p_{\theta}(t)=\mb{p}_{\text{act}}^{(t)}\bm{\eta}^{(t)}+\mb{p}_{\text{bk}}^{(t)}\bm{\nu}^{(t)},\ \forall t>1.
\end{equation}

\subsection{Computational Complexity}
The size $|\mc{S}|$ of the state space is
\begin{equation}
    |\mc{S}|=
    \sum_{a=0}^N\sum_{c=0}^{N-a}(N{-}a{-}c{+}1) 
    =\frac{(N{+}1)(N{+}2)(N{+}3)}{6},
\end{equation}
which is $O\left(N^3\right)$. To compute $\mb{M}$, we need to consider possible activation, which requires $O\left(N^5\right)$ operations. However, there is a more complex operation: computing $\bm{\pi}$ with the eigenvalue method requires $O\left(|\mc{S}|^3\right)$ operations in practice, i.e., $O\left(N^9\right)$. Finally, computing the Peak \gls{aoii} \gls{pmf} up to latency $t$ requires $t$ multiplication of the state distribution by the transition matrix, each of which requires $O\left(|\mc{S}|^2\right)$ operations. The complete calculation of the \gls{paoii} \gls{pmf} up to $t$ then requires $O\left(\max\left[tN^6,N^9\right]\right)$ operations.

\begin{figure}[t!]\vspace{-0.3cm}
\begin{table}[H]
\centering
	\caption{Simulation settings.}
	\label{tab:params}
    \scriptsize
	\begin{tabular}[c]{ccc}
		\toprule
		Parameter & Meaning & Value\\ \midrule
        $N$ & Number of nodes & $20$\\
        $L$ & Payload length & $32$~b\\
        $E$ & Transmission energy & $1$~mJ\\
        $\tau$ & Slot duration & $50$~ms\\
        $\lambda$ & Activation probability & $\{0.01,0.02\}$\\
        $\varepsilon$ & Uplink error probability & $0.1$\\
        $\psi$ & Feedback error probability & $\{0,0.1,0.2\}$\\
		\bottomrule
	\end{tabular}
\end{table}\vspace{-0.6cm}
\end{figure}

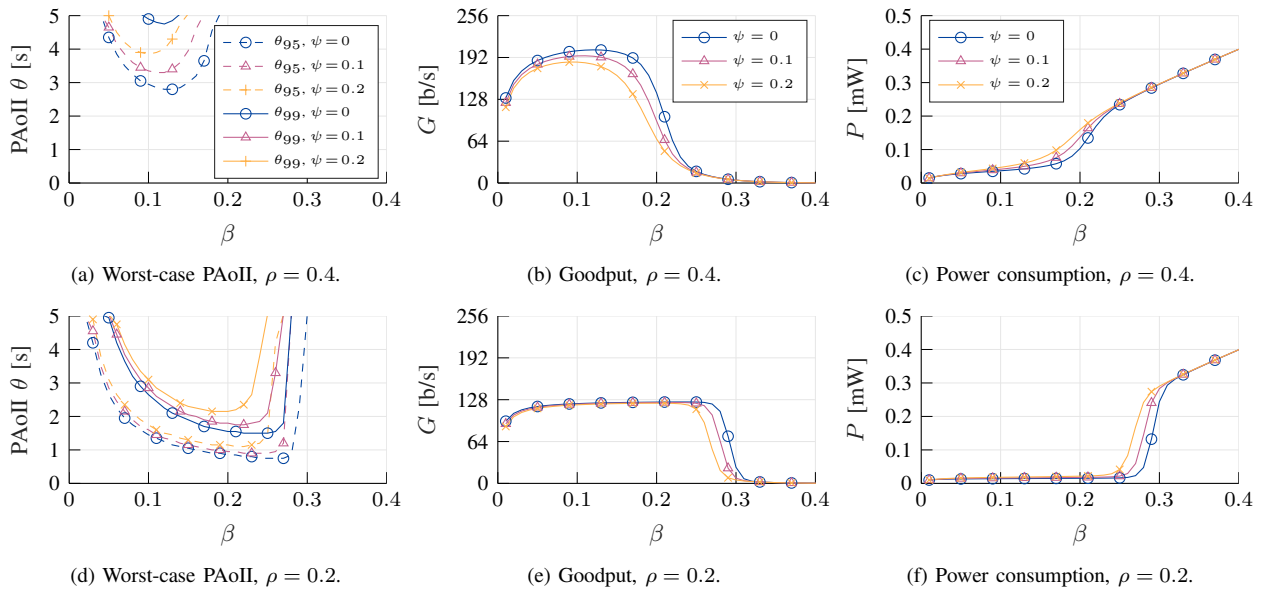
\begin{figure*}[t!]
    \centering
\subfloat[Worst-case PAoII, $\rho=0.4$.\label{fig:aoii_quant_a09_load_40}]
{\begin{tikzpicture}

\begin{axis}[%
width=\sfwidth,
height=\sfheight,
xmin=0,
xmax=0.4,
xlabel style={font={\small\color{white!15!black}}},
xlabel={$\beta$},
ymin=0,
ymax=100,
xmajorgrids,
ymajorgrids,
ytick={0,20,40,60,80,100},
yticklabels={0,1,2,3,4,5},
ylabel style={font={\small\color{white!15!black}}},
ylabel={PAoII $\theta$ [s]},
axis background/.style={fill=white},
axis x line*=bottom,
axis y line*=left,
legend style={draw=white!15!black,    legend columns=1, font=\tiny, at={(0.975, 0.025)}, anchor=south east}
]
\addplot [color=color0, dashed, mark=o, mark options={solid}, mark repeat=4]
  table{./fig/data/quant_95_alpha_09_load_40_eps_10_psi_0_betas.dat};
\addlegendentry{$\theta_{95}$, $\psi\!=\!0$}

\addplot [color=color2, dashed, mark=triangle, mark options={solid}, mark repeat=4]
  table{./fig/data/quant_95_alpha_09_load_40_eps_10_psi_10_betas.dat};
\addlegendentry{$\theta_{95}$, $\psi\!=\!0.1$}

\addplot [color=color4, dashed, mark=+, mark options={solid}, mark repeat=4]
  table{./fig/data/quant_95_alpha_09_load_40_eps_10_psi_20_betas.dat};
\addlegendentry{$\theta_{95}$, $\psi\!=\!0.2$}

\addplot [color=color0, mark=o, mark options={solid}, mark repeat=4]
  table{./fig/data/quant_99_alpha_09_load_40_eps_10_psi_0_betas.dat};
\addlegendentry{$\theta_{99}$, $\psi\!=\!0$}

\addplot [color=color2, mark=triangle, mark options={solid}, mark repeat=4]
  table{./fig/data/quant_99_alpha_09_load_40_eps_10_psi_10_betas.dat};
\addlegendentry{$\theta_{99}$, $\psi\!=\!0.1$}

\addplot [color=color4, mark=+, mark options={solid}, mark repeat=4]
  table{./fig/data/quant_99_alpha_09_load_40_eps_10_psi_20_betas.dat};
\addlegendentry{$\theta_{99}$, $\psi\!=\!0.2$}

\end{axis}
\end{tikzpicture}%
}
\subfloat[Goodput, $\rho=0.4$.\label{fig:goodput_a09_load_40}]
{\begin{tikzpicture}

\begin{axis}[%
width=\sfwidth,
height=\sfheight,
xmin=0,
xmax=0.4,
xlabel style={font={\small\color{white!15!black}}},
xlabel={$\beta$},
ymin=0,
ymax=0.4,
xmajorgrids,
ymajorgrids,
ytick={0,0.1,0.2,0.3,0.4},
yticklabels={0,64,128,192,256},
ylabel style={font={\small\color{white!15!black}}},
ylabel={$G$ [b/s]},
axis background/.style={fill=white},
axis x line*=bottom,
axis y line*=left,
legend style={draw=white!15!black,  legend columns=1, font=\tiny, at={(0.975, 0.975)}, anchor=north east}
]
\addplot [color=color0, mark=o, mark options={solid}, mark repeat=4]
  table[x=beta,y=e0]{./fig/data/goodput_alpha_09_load_40_eps_10_betas.dat};
\addlegendentry{$\psi=0$};
\addplot [color=color2, mark=triangle, mark options={solid}, mark repeat=4]
  table[x=beta,y=e10]{./fig/data/goodput_alpha_09_load_40_eps_10_betas.dat};
\addlegendentry{$\psi=0.1$};

\addplot [color=color4, mark=x, mark options={solid}, mark repeat=4]
  table[x=beta,y=e20]{./fig/data/goodput_alpha_09_load_40_eps_10_betas.dat};
\addlegendentry{$\psi=0.2$};

\end{axis}
\end{tikzpicture}%
}
\subfloat[Power consumption, $\rho=0.4$.\label{fig:energy_a09_load_40}]
{\begin{tikzpicture}

\begin{axis}[%
width=\sfwidth,
height=\sfheight,
xmin=0,
xmax=0.4,
xlabel style={font={\small\color{white!15!black}}},
xlabel={$\beta$},
ymin=0,
ymax=10,
xmajorgrids,
ytick={0,2,4,6,8,10},
yticklabels={0,0.1,0.2,0.3,0.4,0.5},
ymajorgrids,
ylabel style={font={\small\color{white!15!black}}},
ylabel={$P$ [mW]},
axis background/.style={fill=white},
axis x line*=bottom,
axis y line*=left,
legend style={draw=white!15!black,  legend columns=1, font=\tiny, at={(0.025, 0.975)}, anchor=north west}
]
\addplot [color=color0, mark=o, mark options={solid}, mark repeat=4]
  table[x=beta,y=e0]{./fig/data/energy_alpha_09_load_40_eps_10_betas.dat};
\addlegendentry{$\psi=0$};

\addplot [color=color2, mark=triangle, mark options={solid}, mark repeat=4]
  table[x=beta,y=e10]{./fig/data/energy_alpha_09_load_40_eps_10_betas.dat};
\addlegendentry{$\psi=0.1$};

\addplot [color=color4, mark=x, mark options={solid}, mark repeat=4]
  table[x=beta,y=e20]{./fig/data/energy_alpha_09_load_40_eps_10_betas.dat};
\addlegendentry{$\psi=0.2$};

\end{axis}
\end{tikzpicture}%
}\\ \vspace{-0.2cm}
\subfloat[Worst-case PAoII, $\rho=0.2$.\label{fig:aoii_quant_a09_load_20}]
{\begin{tikzpicture}

\begin{axis}[%
width=\sfwidth,
height=\sfheight,
xmin=0,
xmax=0.4,
xlabel style={font={\small\color{white!15!black}}},
xlabel={$\beta$},
ymin=0,
ymax=100,
xmajorgrids,
ymajorgrids,
ytick={0,20,40,60,80,100},
yticklabels={0,1,2,3,4,5},
ylabel style={font={\small\color{white!15!black}}},
ylabel={PAoII $\theta$ [s]},
axis background/.style={fill=white},
axis x line*=bottom,
axis y line*=left
]
\addplot [color=color0, dashed, mark=o, mark options={solid}, mark repeat=4]
  table{./fig/data/quant_95_alpha_09_load_20_eps_10_psi_0_betas.dat};

\addplot [color=color2, dashed, mark=triangle, mark options={solid}, mark repeat=4]
  table{./fig/data/quant_95_alpha_09_load_20_eps_10_psi_10_betas.dat};

\addplot [color=color4, dashed, mark=x, mark options={solid}, mark repeat=4]
  table{./fig/data/quant_95_alpha_09_load_20_eps_10_psi_20_betas.dat};

\addplot [color=color0, mark=o, mark options={solid}, mark repeat=4]
  table{./fig/data/quant_99_alpha_09_load_20_eps_10_psi_0_betas.dat};

\addplot [color=color2, mark=triangle, mark options={solid}, mark repeat=4]
  table{./fig/data/quant_99_alpha_09_load_20_eps_10_psi_10_betas.dat};

\addplot [color=color4, mark=x, mark options={solid}, mark repeat=4]
  table{./fig/data/quant_99_alpha_09_load_20_eps_10_psi_20_betas.dat};

\end{axis}
\end{tikzpicture}%
}
\subfloat[Goodput, $\rho=0.2$.\label{fig:goodput_a09_load_20}]
{\begin{tikzpicture}

\begin{axis}[%
width=\sfwidth,
height=\sfheight,
xmin=0,
xmax=0.4,
xlabel style={font={\small\color{white!15!black}}},
xlabel={$\beta$},
ymin=0,
ymax=0.4,
ytick={0,0.1,0.2,0.3,0.4},
yticklabels={0,64,128,192,256},
xmajorgrids,
ymajorgrids,
ylabel style={font={\small\color{white!15!black}}},
ylabel={$G$ [b/s]},
axis background/.style={fill=white},
axis x line*=bottom,
axis y line*=left
]
\addplot [color=color0, mark=o, mark options={solid}, mark repeat=4]
  table[x=beta,y=e0]{./fig/data/goodput_alpha_09_load_20_eps_10_betas.dat};

\addplot [color=color2, mark=triangle, mark options={solid}, mark repeat=4]
  table[x=beta,y=e10]{./fig/data/goodput_alpha_09_load_20_eps_10_betas.dat};

\addplot [color=color4, mark=x, mark options={solid}, mark repeat=4]
  table[x=beta,y=e20]{./fig/data/goodput_alpha_09_load_20_eps_10_betas.dat};

\end{axis}
\end{tikzpicture}%
}
\subfloat[Power consumption, $\rho=0.2$.\label{fig:energy_a09_load_20}]
{\begin{tikzpicture}

\begin{axis}[%
width=\sfwidth,
height=\sfheight,
xmin=0,
xmax=0.4,
xlabel style={font={\small\color{white!15!black}}},
xlabel={$\beta$},
ymin=0,
ymax=10,
xmajorgrids,
ymajorgrids,
ylabel style={font={\small\color{white!15!black}}},
ytick={0,2,4,6,8,10},
yticklabels={0,0.1,0.2,0.3,0.4,0.5},
ylabel={$P$ [mW]},
axis background/.style={fill=white},
axis x line*=bottom,
axis y line*=left
]
\addplot [color=color0, mark=o, mark options={solid}, mark repeat=4]
  table[x=beta,y=e0]{./fig/data/energy_alpha_09_load_20_eps_10_betas.dat};

\addplot [color=color2, mark=triangle, mark options={solid}, mark repeat=4]
  table[x=beta,y=e10]{./fig/data/energy_alpha_09_load_20_eps_10_betas.dat};

\addplot [color=color4, mark=x, mark options={solid}, mark repeat=4]
  table[x=beta,y=e20]{./fig/data/energy_alpha_09_load_20_eps_10_betas.dat};

\end{axis}
\end{tikzpicture}%
}
    \caption{Performance of the system as a function of $\beta$, with $\alpha=0.9$.}\vspace{-0.3cm}
    \label{fig:aoii_quantiles}
\end{figure*}
\section{Simulation Settings and Results}\label{sec:results}

In order to evaluate the performance of the protocol in different conditions, as well as the possible optimization of its parameters, we first verified the correctness of the results through extensive Monte Carlo simulations. However, for the sake of brevity, the results of this check are not reported in the figures: the Dvoretzky–Kiefer–Wolfowitz inequality~\cite{dvoretzky1956asymptotic} states that the difference between the \gls{cdf} $F(x)$ and the empirical \gls{cdf} $F_L(x)$, resulting from $L$ independent samples, is bounded by:
\begin{equation}
 \Prb{\sup_{x\in\mathbb{R}}(F_L(x){-}F(x)) > \mu} \!\!\leq e^{-2L\mu^2}\!\!,\ \forall\mu \geq \!\sqrt{\frac{\ln(2)}{2L}}.
\end{equation}
Considering $L=10^7$ and $\mu=10^{-3}$, the probability bound is $2\times10^{-9}$. Since the empirical distribution is within this bound at all times, we can conclude that the analysis is correct, and only report the optimization results. The main simulation settings are reported in Table~\ref{tab:params}: the slot time and transmission energy values are consistent with LoRaWAN Class A devices~\cite{bouguera2018energy} using spreading factor $7$.

We first consider the optimization of the backoff probability $\beta$, considering $\alpha=0.9$, i.e., a high probability of an immediate first transmission attempt after an anomaly is detected: the results in terms of the worst-case \gls{paoii}, goodput, and power consumption are shown in Fig.~\ref{fig:aoii_quantiles}. We considered two different levels of the load $\rho=N\lambda$: for $\rho=0.4$, the maximum \gls{aoii} before an anomaly is reported reaches $3$~s in $5\%$ of cases even if the feedback channel is perfect, and is close to $4$ if there is a significant amount of errors on the channel. In general, low values of $\beta$ tend to optimize the worst-case \gls{paoii}, as shown in Fig.~\ref{fig:aoii_quant_a09_load_40}, and considering higher percentiles leads to more conservative choices that reduce congestion. On the other hand, the values that optimize the goodput, shown in Fig.~\ref{fig:goodput_a09_load_40}, tend to be slightly higher. Finally, we can note two inflection points in the power consumption, as shown by Fig.~\ref{fig:energy_a09_load_40}: the first is close to the optimum for goodput, after which power consumption grows significantly with $\beta$. The second inflection point corresponds to a very low goodput, and power consumption grows at a slower pace after it. In these conditions, almost all nodes are in backoff mode all the time, and the growth in the power consumption is given by $\beta$ rather than by newly activated nodes, which use the higher transmission probability $\alpha$. The general trends are slightly different in Fig.~\ref{fig:aoii_quant_a09_load_20}, which shows the higher percentiles of the \gls{paoii} with $\rho=0.2$: the worst-case \gls{paoii} decreases slowly as $\beta$ increases, reaching a minimum just below the point where congestion becomes catastrophic and the goodput sharply drops to $0$. Fig.~\ref{fig:goodput_a09_load_20} shows that goodput is almost flat until that point, but sharply goes to $0$ afterward, and the same two inflection points are visible in the power consumption curve, with a much sharper transition, as Fig.~\ref{fig:energy_a09_load_20} shows.

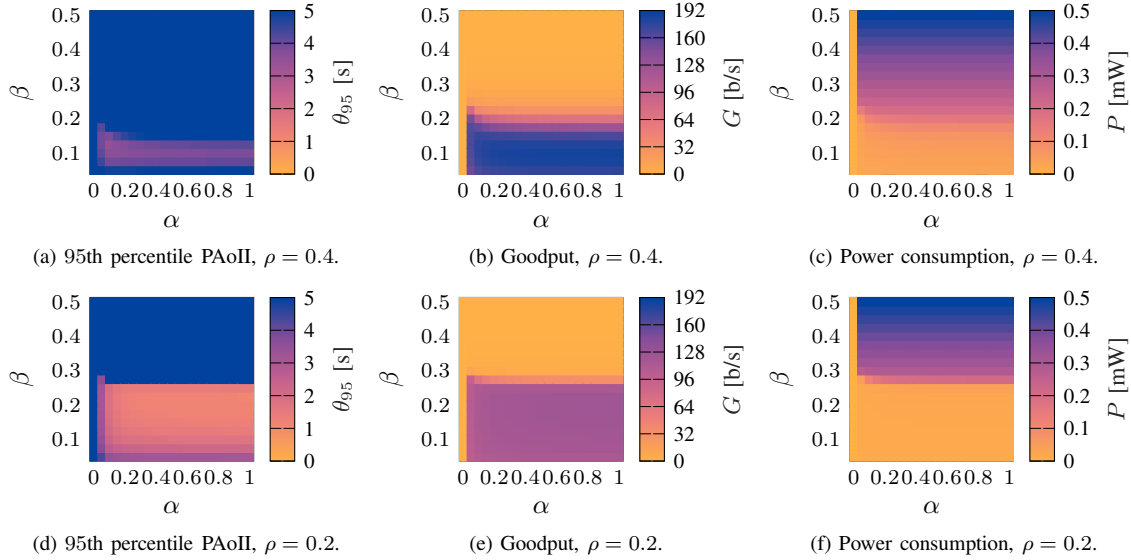
\begin{figure*}[t!]
    \centering
\subfloat[$95$th percentile PAoII, $\rho=0.4$.\label{fig:aoii_heat_eps_20_load_40}]
{\begin{tikzpicture}
    \begin{axis}[
    width=\boxside,
    height=\boxside,
    scale only axis,
    name=lin,
    xlabel=$\alpha$,
    ylabel=$\beta$,
    mesh/cols=19,
    mesh/rows=21,
    xmin=-0.025,
    xmax=1.025,
    ymin=0.0375,
    ymax=0.5125,
    ytick={0.1,0.2,0.3,0.4,0.5},
    xtick={0,0.2,0.4,0.6,0.8,1},
    point meta min=0,
    point meta max=100,
colormap={mymap}{[1pt]
rgb(0pt)=(1, 0.690196078431373, 0.278431372549020);
rgb(1pt)=(0.937254901960784, 0.501960784313726, 0.470588235294118);
rgb(2pt)=(0.764705882352941, 0.376470588235294, 0.556862745098039);
rgb(3pt)=(0.517647058823530, 0.301960784313725, 0.600000000000000);
rgb(4pt)=(0, 0.258823529411765, 0.615686274509804);
 },
    colorbar right,
    colorbar style={
    at={(1.1,1)},
    width=0.15*\pgfkeysvalueof{/pgfplots/parent axis width},
ylabel style={font={\small\color{white!15!black}}},ylabel={$\theta_{95}$ [s]},
                    ytick={0,20,...,100},
                    yticklabels={0,1,...,5}}
]
    \addplot[matrix plot*, point meta=explicit] file {./fig/data/heatmap_aoii_rho_40_eps_10_psi_20.dat};
\end{axis}

\end{tikzpicture}}
\subfloat[Goodput, $\rho=0.4$.\label{fig:goodput_heat_eps_20_load_40}]
{\begin{tikzpicture}
    \begin{axis}[
    width=\boxside,
    height=\boxside,
    scale only axis,
    name=lin,
    xlabel=$\alpha$,
    ylabel=$\beta$,
    mesh/cols=19,
    mesh/rows=21,
    xmin=-0.025,
    xmax=1.025,
    ymin=0.0375,
    ymax=0.5125,
    ytick={0.1,0.2,0.3,0.4,0.5},
    xtick={0,0.2,0.4,0.6,0.8,1},
    point meta min=0,
    point meta max=0.3,
colormap={mymap}{[1pt]
rgb(0pt)=(1, 0.690196078431373, 0.278431372549020);
rgb(1pt)=(0.937254901960784, 0.501960784313726, 0.470588235294118);
rgb(2pt)=(0.764705882352941, 0.376470588235294, 0.556862745098039);
rgb(3pt)=(0.517647058823530, 0.301960784313725, 0.600000000000000);
rgb(4pt)=(0, 0.258823529411765, 0.615686274509804);
 },
    colorbar right,
    colorbar style={
    at={(1.1,1)},
    width=0.15*\pgfkeysvalueof{/pgfplots/parent axis width},
ylabel style={font={\small\color{white!15!black}}},ylabel={$G$ [b/s]},
                    yticklabel style={
                        /pgf/number format/fixed,
                        /pgf/number format/precision=2
                },
                scaled y ticks=false,
                    ytick={0,0.05,...,0.3},
                    yticklabels={0,32,...,384}}
]
    \addplot[matrix plot*, point meta=explicit] file {./fig/data/heatmap_goodput_rho_40_eps_10_psi_20.dat};
\end{axis}

\end{tikzpicture}}
\subfloat[Power consumption, $\rho=0.4$.\label{fig:power_heat_eps_20_load_40}]
{\begin{tikzpicture}
    \begin{axis}[
    width=\boxside,
    height=\boxside,
    scale only axis,
    name=lin,
    xlabel=$\alpha$,
    ylabel=$\beta$,
    mesh/cols=19,
    mesh/rows=21,
    xmin=-0.025,
    xmax=1.025,
    ymin=0.0375,
    ymax=0.5125,
    ytick={0.1,0.2,0.3,0.4,0.5},
    xtick={0,0.2,0.4,0.6,0.8,1},
    point meta min=0,
    point meta max=10,
colormap={mymap}{[1pt]
rgb(0pt)=(1, 0.690196078431373, 0.278431372549020);
rgb(1pt)=(0.937254901960784, 0.501960784313726, 0.470588235294118);
rgb(2pt)=(0.764705882352941, 0.376470588235294, 0.556862745098039);
rgb(3pt)=(0.517647058823530, 0.301960784313725, 0.600000000000000);
rgb(4pt)=(0, 0.258823529411765, 0.615686274509804);
 },
    colorbar right,
    colorbar style={
    at={(1.1,1)},
    width=0.15*\pgfkeysvalueof{/pgfplots/parent axis width},
ylabel style={font={\small\color{white!15!black}}},ylabel={$P$ [mW]},
                    ytick={0,2,...,10},
                    yticklabels={0,0.1,0.2,0.3,0.4,0.5}}
]
    \addplot[matrix plot*, point meta=explicit] file {./fig/data/heatmap_power_rho_40_eps_10_psi_20.dat};
\end{axis}

\end{tikzpicture}}\\ \vspace{-0.2cm}
\subfloat[$95$th percentile PAoII, $\rho=0.2$.\label{fig:aoii_heat_eps_20_load_20}]
{\begin{tikzpicture}
    \begin{axis}[
    width=\boxside,
    height=\boxside,
    scale only axis,
    name=lin,
    xlabel=$\alpha$,
    ylabel=$\beta$,
    mesh/cols=19,
    mesh/rows=21,
    xmin=-0.025,
    xmax=1.025,
    ymin=0.0375,
    ymax=0.5125,
    ytick={0.1,0.2,0.3,0.4,0.5},
    xtick={0,0.2,0.4,0.6,0.8,1},
    point meta min=0,
    point meta max=100,
colormap={mymap}{[1pt]
rgb(0pt)=(1, 0.690196078431373, 0.278431372549020);
rgb(1pt)=(0.937254901960784, 0.501960784313726, 0.470588235294118);
rgb(2pt)=(0.764705882352941, 0.376470588235294, 0.556862745098039);
rgb(3pt)=(0.517647058823530, 0.301960784313725, 0.600000000000000);
rgb(4pt)=(0, 0.258823529411765, 0.615686274509804);
 },
    colorbar right,
    colorbar style={
    at={(1.1,1)},
    width=0.15*\pgfkeysvalueof{/pgfplots/parent axis width},
ylabel style={font={\small\color{white!15!black}}},ylabel={$\theta_{95}$ [s]},
                    ytick={0,20,...,100},
                    yticklabels={0,1,...,5}}
]
    \addplot[matrix plot*, point meta=explicit] file{./fig/data/heatmap_aoii_rho_20_eps_10_psi_20.dat};
\end{axis}

\end{tikzpicture}}
\subfloat[Goodput, $\rho=0.2$.\label{fig:goodput_heat_eps_20_load_20}]
{\begin{tikzpicture}
    \begin{axis}[
    width=\boxside,
    height=\boxside,
    scale only axis,
    name=lin,
    xlabel=$\alpha$,
    ylabel=$\beta$,
    mesh/cols=19,
    mesh/rows=21,
    xmin=-0.025,
    xmax=1.025,
    ymin=0.0375,
    ymax=0.5125,
    ytick={0.1,0.2,0.3,0.4,0.5},
    xtick={0,0.2,0.4,0.6,0.8,1},
    point meta min=0,
    point meta max=0.3,
colormap={mymap}{[1pt]
rgb(0pt)=(1, 0.690196078431373, 0.278431372549020);
rgb(1pt)=(0.937254901960784, 0.501960784313726, 0.470588235294118);
rgb(2pt)=(0.764705882352941, 0.376470588235294, 0.556862745098039);
rgb(3pt)=(0.517647058823530, 0.301960784313725, 0.600000000000000);
rgb(4pt)=(0, 0.258823529411765, 0.615686274509804);
 },
    colorbar right,
    colorbar style={
    at={(1.1,1)},
    width=0.15*\pgfkeysvalueof{/pgfplots/parent axis width},
ylabel style={font={\small\color{white!15!black}}},ylabel={$G$ [b/s]},
                    yticklabel style={
                        /pgf/number format/fixed,
                        /pgf/number format/precision=2
                },
                scaled y ticks=false,
                    ytick={0,0.05,...,0.3},
                    yticklabels={0,32,...,384}}
]
    \addplot[matrix plot*, point meta=explicit] file {./fig/data/heatmap_goodput_rho_20_eps_10_psi_20.dat};
\end{axis}

\end{tikzpicture}}
\subfloat[Power consumption, $\rho=0.2$.\label{fig:power_heat_eps_20_load_20}]
{\begin{tikzpicture}
    \begin{axis}[
    width=\boxside,
    height=\boxside,
    scale only axis,
    name=lin,
    xlabel=$\alpha$,
    ylabel=$\beta$,
    mesh/cols=19,
    mesh/rows=21,
    xmin=-0.025,
    xmax=1.025,
    ymin=0.0375,
    ymax=0.5125,
    ytick={0.1,0.2,0.3,0.4,0.5},
    xtick={0,0.2,0.4,0.6,0.8,1},
    point meta min=0,
    point meta max=10,
colormap={mymap}{[1pt]
rgb(0pt)=(1, 0.690196078431373, 0.278431372549020);
rgb(1pt)=(0.937254901960784, 0.501960784313726, 0.470588235294118);
rgb(2pt)=(0.764705882352941, 0.376470588235294, 0.556862745098039);
rgb(3pt)=(0.517647058823530, 0.301960784313725, 0.600000000000000);
rgb(4pt)=(0, 0.258823529411765, 0.615686274509804);
 },
    colorbar right,
    colorbar style={
    at={(1.1,1)},
    width=0.15*\pgfkeysvalueof{/pgfplots/parent axis width},
ylabel style={font={\small\color{white!15!black}}},ylabel={$P$ [mW]},
                    ytick={0,2,...,10},
                    yticklabels={0,0.1,0.2,0.3,0.4,0.5}}
]
    \addplot[matrix plot*, point meta=explicit] file {./fig/data/heatmap_power_rho_20_eps_10_psi_20.dat};
\end{axis}

\end{tikzpicture}}
    \caption{Performance of the system as a function of $\alpha$ and $\beta$, with $\psi=0.2$.}\vspace{-0.3cm}
    \label{fig:heatmaps}
\end{figure*}

We can extend the analysis to optimize both $\alpha$ and $\beta$, as shown in Fig.~\ref{fig:heatmaps}. The region with the lowest worst-case \gls{paoii} also results in a solid performance in terms of both goodput and energy consumption. The key parameter in the analysis is $\beta$: reducing the value of $\alpha$ may increase the average \gls{aoii}, but does not have a significant impact on $95$th percentile performance unless it becomes very low, whereas $\beta$ has a stronger impact. The same effect is visible in terms of goodput and power consumption: Figs.~\ref{fig:power_heat_eps_20_load_40} and \ref{fig:power_heat_eps_20_load_20} confirm it, showing a negligible effect of $\alpha$ on power consumption. 

This phenomenon is caused by the disproportionate effect of a sensor's behavior after a collision on other nodes, while the impact of the first transmission attempt is already mitigated by the relatively low activation probability for each sensor, so that collisions often involve nodes that are already in backoff mode. Setting a relatively low value of $\beta$, which depends on the system load, can help prevent these cascades of collisions, while still allowing nodes to report anomalies relatively often.

\section{Conclusions}\label{sec:conc}

This work presents a closed-form analysis of the \gls{paoii} in a reactive slotted ALOHA system for reporting anomalies, highlighting the effect of imperfect feedback on protocol performance. Our analysis finds that aiming for a high reliability in terms of \gls{paoii}, i.e., considering high percentiles of the distribution as an optimization target, leads to more stable configurations, which also have benefits in terms of goodput and energy efficiency. On the other hand, even considering the $95$th percentile as an optimization target leads the system right to the edge of instability, with significant consequences for the application if the statistics of the anomaly generation process change over time. Furthermore, we find that adapting the probability of transmission for nodes that detect a new anomaly has a negligible effect on both \gls{paoii} and more traditional metrics, while the backoff probability becomes a crucial parameter in the optimization.

Considering an imperfect feedback channel changes the outcomes of the optimization significantly, and further research on the subject is required to obtain a closed-form optimum for a given configuration; studying the robustness and adaptability of the protocol to changes in the system statistics is another key research direction that can exploit the analytical results we derived in this paper. 

\section*{Acknowledgments}
This project was partly funded under the Italian National Recovery and Resilience Plan (NRRP), as part of the RESTART partnership (PE0000001) and the ``Young Researchers'' grant REDIAL (SoE0000009), under the European Union NextGenerationEU Project. A. Munari acknowledges the financial support by the Federal Ministry of Education and Research of Germany in the programme of ”Souver\"an. Digital. Vernetzt.” Joint project 6G-RIC, project identification number: 16KISK022. P. Popovski's work is supported by the Villum Investigator grant ``WATER'' from the Velux Foundation.

\bibliographystyle{IEEEtran}
\bibliography{IEEEabrv,biblio.bib}

\end{document}